\documentclass[seceq]{ptptex}

\notypesetlogo                       

\title{
Position Uncertainty in the Heisenberg Uncertainty Relation
}

\author{
Seiji \textsc{Kosugi}%
}

\inst{
Department of Food and Nutrition, Shukutoku Junior College, 6-36-4 Maenocho, Itabashi-ku, Tokyo 174-8631, Japan
}

\abst{
Position measurements are examined under the assumption that object position $\hat{x}_t$ and probe position $\hat{X}_t$ just after the measurement are expressed by a linear combination of positions $\hat{x}_0$ and $\hat{X}_0$ just before the measurement.
The Heisenberg uncertainty relation between the position uncertainty and momentum disturbance holds when the measurement error $\epsilon(x_t)$ for the object position $\hat{x}_t$ is adopted as the position uncertainty.
However, the uncertainty in the measurement result obtained for $\hat{x}_0$ is the standard deviation of the measurement result, and not the measurement error $\epsilon(x_0)$.
This difference is due to the reduction of a wave packet.
The validity of the linearity assumption is examined in detail.

}

\PTPindex{060, 064}  

\begin{document}
\maketitle

\section{Introduction}

In several texts, the Heisenberg uncertainty relation for position and momentum is expressed as
\begin{eqnarray}
\sigma(q)\sigma(p) \geq \hbar/2,
\end{eqnarray}
where $\sigma(q)$ and $\sigma(p)$ are the standard deviations of position $\hat{q}$ and  momentum $\hat{p}$, respectively, in any quantum state.
This relation can be easily derived by using the commutation relation $[\, \hat{q},\, \hat{p} \, ]=i\hbar$. 
However, there exists another well-known relation that was demonstrated by Heisenberg in his thought experiment using a $\gamma$-ray microscope:\cite{WHZP,WHPP}
\begin{equation}
   (\Delta q)_{\rm m} (\Delta p)_{\rm dis} \geq \hbar/2,
\end{equation}
where $(\Delta q)_{\rm m}$ is the uncertainty in a measured position and $(\Delta p)_{\rm dis}$ is the momentum disturbance in the object caused by the position measurement.
The validity of Eq. (1$\cdot$2) has been criticized, and a general proof for this relation has not yet been determined.

As noted by Braginsky and Khalili,\cite{BrKhQM} \ 
although Eqs. (1$\cdot$1) and (1$\cdot$2) are similar in form, their physical meanings are different.
Equation (1$\cdot$1) expresses the fundamental property of the physical state of a quantum object.
The quantity $\sigma(q)$ expresses the statistical fluctuation of the object position. The same also holds true for $\sigma(p)$ with regard to the object momentum.
However, Eq. (1$\cdot$2) is associated with the measurement process.
The position uncertainty $ (\Delta q)_{\rm m} $ is the uncertainty in the position-measurement result, and the momentum uncertainty $(\Delta p)_{\rm dis}$ is the disturbance in the object caused by the measurement process. Nevertheless, both equations are considered to express the Heisenberg uncertainty relation.

Ozawa proposed a certain measurement that violates uncertainty relation (1$\cdot$2) between the measurement error for the pre-measurement position $\hat{x}_0$ and the momentum disturbance.\cite{MOPL1,MOPL2,MOPR} \
Furthermore, he reformulated the uncertainty relation and obtained its generalization, which was proven to be valid for an arbitrary measurement. 
However, there are several questionable points in the measurement errors that he defined.
Its physical meaning is not clear enough, as pointed out by Koshino and Shimizu.\cite{KSPR} \ 
Ozawa considered the position uncertainty in uncertainty relation (1$\cdot$2) to be the measurement error $\epsilon(x_0)$ for the object position $\hat{x}_0$ just before the measurement.
However, this measurement error cannot be regarded as the position uncertainty in uncertainty relation (1$\cdot$2).
For example, when one obtains a certain value by a one-time measurement of the position $\hat{x}_0$ with $\epsilon(x_0)=0$, then it does not mean that it certainly has the value before the measurement.
In this case, its result has the uncertainty of the initial spreading $\sigma(x_0)$ of the object probability density.
The uncertainty in the measurement result for the object position $\hat{x}_0$ is the standard deviation of the measurement result and not the measurement error $\epsilon(x_0)$.

In this paper, we reexamine the Heisenberg uncertainty relation given by Eq. (1$\cdot$2).
Under the linearity assumption that the object position $\hat{x}_t$ and probe position $\hat{X}_t$ just after the measurement are expressed by the linear combination of the positions $\hat{x}_0$ and $\hat{X}_0$ just before the measurement, we will prove that the Heisenberg uncertainty relation between the position uncertainty and momentum disturbance holds when the measurement error $\epsilon(x_t)$ for the object position $\hat{x}_t$ just after the measurement is adopted as the position uncertainty.
This type of uncertainty relation holds also for the interaction proposed by Ozawa in order to demonstrate that there exists a measurement to violate uncertainty relation (1$\cdot$2).
However, the uncertainty relation does not hold when the measurement error $\epsilon(x_0)$ for the pre-measurement position $\hat{x}_0$ is adopted.
The uncertainty in the measurement result obtained for the pre-measurement position $\hat{x}_0$ is the standard deviation of the measurement result.
When this uncertainty is adopted, the uncertainty relation is valid.
The validity of the measurement errors proposed by Ozawa for the object positions $\hat{x}_0$ and $\hat{x}_t$ is examined.
The justification for the linearity assumption for the position operators $\hat{x}_t$ and $\hat{X}_t$ is discussed in detail.

\section{Measurement model}

We consider the measurement of the positions $\hat{x}_0$ and $\hat{x}_t$ of a one-dimensional microscopic object.
It is assumed that the object and probe, which is a part of the apparatus, interact in the time interval $(0,t)$. 
After the interaction of the object with the probe, 
the probe position $\hat{X}_t$ is measured using another measurement apparatus that is coupled only with the probe, and by using this value, the measurement results of the object positions $\hat{x}_0$ and $\hat{x}_t$ are obtained.
It is assumed that the probe position $\hat{X}_t$ can be precisely measured without further disturbing the object momentum.

Let $\hat{U}$ be a unitary operator representing the time evolution of the object and probe for the time interval $(0,t)$.
Then, the object position $\hat{x}_t$ and probe position $\hat{X}_t$ after the measurement are $\hat{U}^{\dagger} (\hat{x}_0 \otimes \hat{I}) \hat{U}$ and $\hat{U}^{\dagger} (\hat{I} \otimes \hat{X}_0) \hat{U}$, respectively. 
We hereinafter abbreviate the tensor product $\hat{x}_0 \otimes \hat{I}$ as $\hat{x}_0$.

In our measurement model, we estimate the measurement result of the object position $\hat{x}_0$ using the measurement result of the probe position $\hat{X}_t$; therefore, the position $\hat{X}_t$ should be a function of $\hat{x}_0$ and $\hat{X}_0$.
The initial probe state $|\xi_0 \rangle$ is arranged in a prescribed position with a small position fluctuation before the measurement, which corresponds to the zero-position setting of a pointer.

von Neumann introduced the interaction
\begin{equation}
    \hat{H}=K\hat{x}_0\hat{P}_0
\end{equation}
between the object and probe in his famous theory of quantum measurement,\cite{VNMF} \ where $\hat{P}_0$ is the probe momentum before the measurement and $K$ is a coupling constant and assumed to be so strong that the free Hamiltonians of the object and probe can be neglected.
Using the time evolution $\hat{U}=\exp(-i\hat{H}t/\hbar)$ and assuming $g_0 \equiv Kt=1$, Caves \cite{Caves} and Ozawa \cite{MOPL1,MOPL2} solved the Heisenberg equation of motion and obtained
\begin{equation}
   \hat{x}_t=\hat{x}_0, \quad \hat{X}_t= \hat{x}_0+\hat{X}_0.
\end{equation}

Ozawa proposed the following interaction to demonstrate that there exists a measurement that violates uncertainty relation (1$\cdot$2):\cite{MOPL1,MOPL2}
\begin{equation}
   \hat{H}=\frac{K\pi}{3\sqrt{3}}(2\hat{x}_0\hat{P}_0-2\hat{p}_0\hat{X}_0+\hat{x}_0\hat{p}_0-\hat{X}_0\hat{P}_0),
\end{equation}
where $\hat{p}_0$ is the object momentum before the measurement.
For this interaction with $g_0=1$, he obtained 
\begin{equation}
   \hat{x}_t=\hat{x}_0-\hat{X}_0 , \quad \hat{X}_t=\hat{x}_0.
\end{equation}

However, it is not suitable to use the interactions given by Eqs. (2$\cdot$1) and (2$\cdot$3) for estimating the momentum disturbance caused by the position measurement because these interactions with $g_0=1$ violate the law of the conservation of momentum.
Here, we propose the following interaction that conserves the total momentum of the object and probe independently of the value of $g_0$:
\begin{equation}
   \hat{H}=K(\hat{p}_0+\hat{P}_0)(\hat{X}_0-\hat{x}_0).
\end{equation}
For this interaction, we obtain
\begin{equation}
   \hat{x}_t=(1-g_0)\hat{x}_0+g_0\hat{X}_0, \quad \hat{X}_t=-g_0\hat{x}_0+(1+g_0)\hat{X}_0.
\end{equation}

For all the interactions mentioned above, the position operators $\hat{x}_t$ and $\hat{X}_t$ are expressed by the linear combination of $\hat{x}_0$ and $\hat{X}_0$:
\begin{eqnarray}
   \hat{x}_t=\alpha_1 \hat{x}_0 +\alpha_2 \hat{X}_0, \quad
   \hat{X}_t=\beta_1 \hat{x}_0 + \beta_2  \hat{X}_0,
\end{eqnarray}
where $\alpha_1, \alpha_2, \beta_1$, and $\beta_2$ are real numbers.
It is easily proven that when the interaction conserves the total momentum, we obtain 
\begin{eqnarray}
   \alpha_1 +  \alpha_2 = 1, \quad \beta_1 +  \beta_2 = 1.
\end{eqnarray}
Furthermore, we obtain $|\Gamma | = 1$ from the condition that the evolution $\hat{U}$ is unitary, where $\Gamma \equiv 1/(\alpha_1 \beta_2 - \alpha_2 \beta_1)$.
Therefore, of four parameters $\alpha_1$, $\alpha_2$, $\beta_1$, and $\beta_2$, only one independent parameter exists.
In this paper, we consider only the position measurements for which relation (2$\cdot$7) holds.
The validity of this assumption will be examined later in $\S$7.

\section{Measurement error}

In our position-measurement model, the measurement result $(x_0)_{\rm exp}$ of the object position $\hat{x}_0$ just before the measurement is estimated by using the measurement result $X$ of the probe position $\hat{X}_t$ just after the measurement.
Following Dirac's method, we let $|x_0\rangle$ ($|X_0\rangle$) and $x_0$ ($X_0$) denote the eigenket and eigenvalue of $\hat{x}_0$ ($\hat{X}_0$), respectively.
When the initial states of the object and probe are eigenstates $|x_0 \rangle$ and $|X_0 \rangle$, respectively, we obtain the relation $X=\beta_1 x_0+\beta_2 X_0$ from Eq. (2$\cdot$7).
In this case, the measurement result $(x_0)_{\rm exp}$ for the pre-measurement position $\hat{x}_0$ of the object should be defined by 
\begin{eqnarray}
   (x_0)_{\rm exp} = \frac{1}{\beta_1} X - \frac{\beta_2}{\beta_1} X_0 \quad (\beta_1 \neq 0).  \nonumber
\end{eqnarray}
When the probe has the spreading of the probability density, we cannot know which position component $X_0$ of the probe wave function interacts with the object.
Therefore, we define the measurement result $(x_0)_{\rm exp}$ by the value of $x_0$ obtained when $X_0$ is set equal to $\langle \hat{X}_0 \rangle $ in the above equation:
\begin{eqnarray}
   (x_0)_{\rm exp} = \frac{1}{\beta_1} X - \frac{\beta_2}{\beta_1} \langle \hat{X}_0 \rangle \quad (\beta_1 \neq 0),
\end{eqnarray}
where $\langle \hat{X}_0 \rangle $ is the average value of the probe position $\hat{X}_0$.
It is clear that the value of $(x_0)_{\rm exp}$ defined above is identical to that of the initial object position $x_0$ in the case where the initial states of the object and  probe are eigenstates $|x_0 \rangle$ and $|X_0 \rangle$, respectively.
When $\beta_1=0$, this apparatus cannot measure the object position $\hat{x}_0$ because it is not expressed as a function of $\hat{X}_t$ and $\hat{X}_0$.
The values of $\langle \hat{X}_0 \rangle $, $\beta_1$, and $\beta_2$, which are the fundamental parameters of the apparatus, are known before the measurement.

We define the operator for the measurement result $(x_0)_{\rm exp}$ by
\begin{eqnarray}
   (\hat{x}_0)_{\rm exp} &=& \frac{1}{\beta_1} \hat{X}_t - \frac{\beta_2}{\beta_1} \langle \hat{X}_0 \rangle \hat{I} \nonumber  \\
   &=&  \hat{x}_0+\frac{\beta_2}{\beta_1}( \hat{X}_0-\langle \hat{X}_0 \rangle \hat{I}).
\end{eqnarray}
Then, for arbitrary initial states of the object and probe, and for an arbitrary value of $\beta_1 \neq 0$, the average value of $(\hat{x}_0)_{\rm exp}$ is identical to
that of the object position $\hat{x}_0$ just before the measurement.
It is reasonable that $\langle (\hat{x}_0)_{\rm exp} \rangle$ coincides with $\langle \hat{x}_0 \rangle$.

Next, we define the measurement error $\epsilon(x_0)$ for the pre-measurement position $\hat{x}_0$ by 
\begin{eqnarray}
   \epsilon(x_0) &=& \langle \phi_0,\xi_0 | \{ (\hat{x}_0)_{\rm exp}-\hat{x}_0 \} ^{2}| \phi_0,\xi_0 \rangle ^{1/2}   \\
   &=& | \frac{\beta_2}{\beta_1} | \sigma(X_0),  \nonumber
\end{eqnarray}
where $|\phi_0\rangle $ and $|\xi_0 \rangle$ are the initial states of the object and probe, respectively, and $\sigma(X_0)$ is the standard deviation of the probe position $\hat{X}_0$ before the measurement.
We represent the tensor product $|\phi_0\rangle \otimes |\xi_0 \rangle$ as $|\phi_0, \xi_0 \rangle$.
Then, in order to understand the physical meaning of the above definition, we express it in the following form:
\begin{eqnarray}
 \epsilon(x_0)^2 = \int \Big\{x_0+\frac{\beta_2}{\beta_1}( X_0-\langle \hat{X}_0 \rangle )-x_0 \Big\}^2 |\phi_0(x_0)|^2 |\xi_0(X_0)|^2 dx_0dX_0. \nonumber
\end{eqnarray}
This indicates that when the object and probe whose states are $|x_0 \rangle$ and $|X_0 \rangle$, respectively, interact, then the measurement result $(x_0)_{\rm exp}$ and true value of the measured observable $\hat{x}_0$ are $x_0+\frac{\beta_2}{\beta_1}( X_0-\langle \hat{X}_0 \rangle )$ and $x_0$, respectively, and that the measurement error defined by Eq. (3$\cdot$3) is certainly the error obtained when the measurement result (3$\cdot$1) is adopted.

Next, we define the measurement result $(x_t)_{\rm exp}$ for the post-measurement position $\hat{x}_t$ using the measurement result $X$ of the probe position as
\begin{eqnarray}
   (x_t)_{\rm exp} &=& \alpha_1 (x_0)_{\rm exp}+\alpha_2 \langle \hat{X}_0 \rangle    \nonumber  \\
               &=& \frac{\alpha_1}{\beta_1} X - \frac{1}{\beta_1 \Gamma} \langle \hat{X}_0 \rangle .
\end{eqnarray}
Clearly, the value $(x_t)_{\rm exp}$ estimated with the above equation is identical to that of the post-measurement position $\hat{x}_t$ (2$\cdot$7), when the initial states of the object and probe are eigenstates $|x_0 \rangle$ and $|X_0 \rangle$, respectively, because then $(x_0)_{\rm exp}=x_0$ and $ \langle \hat{X}_0 \rangle =X_0$.

In the same manner as the measurement error $\epsilon(x_0)$, we obtain the measurement error $\epsilon(x_t)$ for the post-measurement position $\hat{x}_t$:
\begin{eqnarray}
   \epsilon(x_t) &=& \langle \phi_0,\xi_0 | \{ (\hat{x}_t)_{\rm exp}-\hat{x}_t \} ^{2}| \phi_0,\xi_0 \rangle ^{1/2}  \\
                 &=& \frac{1}{ | \beta_1 | }\sigma(X_0),  \nonumber
\end{eqnarray}
where
\begin{eqnarray}
   (\hat{x}_t)_{\rm exp} &=& \frac{\alpha_1}{\beta_1} \hat{X}_t - \frac{1}{\beta_1 \Gamma} \langle \hat{X}_0 \rangle \hat{I}  \\
    &=& \hat{x}_t +\frac{1}{\beta_1 \Gamma}(\hat{X}_0-\langle \hat{X}_0 \rangle \hat{I}).
\end{eqnarray}

It is easily found from Eq. (3$\cdot$7) that the average value of $(\hat{x}_t)_{\rm exp}$ is identical to that of the object position $\hat{x}_t$ just after the measurement for arbitrary initial states of the object and probe, and for an arbitrary value of $\beta_1 \neq 0$.
It should be noted that the measurement error $\epsilon(x_t)$ (3$\cdot$5) differs from the measurement error $\epsilon(x_0)$ (3$\cdot$3).

The measurement error $\epsilon(x_t)$ can also be derived by using the measurement result $(x_t)_{\rm exp}$ in Eq. (3$\cdot$4) and the wave function $\langle x,X|\hat{U}|\phi_0,\xi_0 \rangle $ of the system composed of the object and probe after the measurement.
This corresponds to the representation in Schr\"{o}dinger picture.
Since the probability density to obtain the measurement results $x$ and $X$ for the post-measurement positions  $\hat{x}_t$ and $\hat{X}_t$, respectively, is $|\langle x,X|\hat{U}|\phi_0,\xi_0 \rangle |^2$, the measurement error $\epsilon(x_t)$ is given by 
\begin{eqnarray}
 \epsilon(x_t)^2 &=& \int \Big\{\frac{\alpha_1}{\beta_1}X-\frac{1}{\beta_1 \Gamma}\langle \hat{X}_0 \rangle -x \Big\}^2 |\langle x,X|\hat{U}|\phi_0,\xi_0 \rangle |^2 dx dX  \\
 &=& \langle \phi_0,\xi_0 | \hat{U}^{-1} \Big\{ \frac{\alpha_1}{\beta_1}\hat{X}_0-\frac{1}{\beta_1 \Gamma}\langle \hat{X}_0 \rangle \hat{I} -\hat{x}_0 \Big\} ^{2} \hat{U}| \phi_0,\xi_0 \rangle . 
\end{eqnarray}
By using the following equalities $\hat{U}^{-1} \hat{X}_0 \hat{U}=\hat{X}_t$ and $\hat{U}^{-1} \hat{x}_0 \hat{U}=\hat{x}_t$, it is found that the right-hand side of Eq. (3$\cdot$9) is equal to the square of that of Eq. (3$\cdot$5).

Furthermore, from Eq. (3$\cdot$8), we obtain the square of the measurement error $\epsilon_X (x_t)$ obtained when the measurement result of the position $\hat{X}_t$ is $X$ as follows:
\begin{eqnarray}
 \epsilon_X (x_t)^2 &=& \int \Big\{\frac{\alpha_1}{\beta_1}X-\frac{1}{\beta_1 \Gamma}\langle \hat{X}_0 \rangle -x \Big\}^2 |\langle x,X|\hat{U}|\phi_0,\xi_0 \rangle |^2 dx /P(X),  \\
 P(X) &=& \int |\langle x,X|\hat{U}|\phi_0,\xi_0 \rangle |^2 dx .
\end{eqnarray}
Note that $ P(X)$ is the probability density to obtain the measurement result $X$ for the position $\hat{X}_t$.
From Eq. (3$\cdot$10), we obtain
\begin{eqnarray}
 \epsilon(x_t)^2 = \int \epsilon_X (x_t)^2 P(X) dX, 
\end{eqnarray}
indicating that the error $\epsilon(x_t)$ is an average error, averaged with the weight function $P(X)$ over the possible measurement results of the probe position $\hat{X}_t$.

In the same manner as the measurement error $\epsilon(x_t)$, from Eq. (3$\cdot$1), we obtain the following measurement error $\epsilon_X(x_0)$, which is the measurement error for the pre-measurement position $\hat{x}_0$ obtained when the measurement result of the position $\hat{X}_t$ is $X$:
\begin{eqnarray}
 \epsilon_X (x_0)^2 = \int \Big\{\frac{1}{\beta_1}X-\frac{\beta_2}{\beta_1}\langle \hat{X}_0 \rangle -\Gamma (\beta_2 x-\alpha_2 X) \Big\}^2 |\langle x,X|\hat{U}|\phi_0,\xi_0 \rangle |^2 dx /P(X), \nonumber
\end{eqnarray}
using the equality $\hat{U} \hat{x}_0 \hat{U}^{-1} = \Gamma(\beta_2 \hat{x}_0-\alpha_2 \hat{X}_0)$.
From the above equation, we obtain
\begin{eqnarray} 
\epsilon(x_0)^2 = \int \epsilon_X (x_0)^2 P(X) dX.
\end{eqnarray}

\section{Measurement error by Ozawa}

Ozawa claimed that the uncertainty relation does not hold when the interaction given by Eq. (2$\cdot$3) and the following measurement error for the pre-measurement position $\hat{x}_0$ are adopted:\cite{MOPL1,MOPL2}
\begin{eqnarray}
   \epsilon^{\rm Ozawa}(x_0) = \langle \phi_0,\xi_0 | \{ \hat{X}_t-\hat{x}_0 \} ^{2}| \phi_0,\xi_0 \rangle ^{1/2}.
\end{eqnarray}

As already mentioned, it is not suitable to use interaction (2$\cdot$3) for estimating the momentum disturbance caused by the position measurement.
On comparing the two definitions (3$\cdot$3) and (4$\cdot$1) of the measurement error, it is evident  that Ozawa adopted the measurement result $X$ of the probe position $\hat{X}_t$ as the measurement result $(x_0)_{\rm exp}$ of the object position $\hat{x}_0$:
\begin{eqnarray}
   (\hat{x}_0)_{\rm exp} ^{\rm Ozawa} \equiv \hat{X}_t.
\end{eqnarray}

A measurement theory should correctly reproduce the Born rule of probability for any state $| \phi_0 \rangle$, when the measurement error $\epsilon(x_0)$ is zero.
Then, the following condition should be fulfilled for any $| \phi_0 \rangle$:
\begin{eqnarray}
   \langle \phi_0,\xi_0 | (\hat{x}_0)_{\rm exp} | \phi_0,\xi_0 \rangle =    \langle \phi_0| \hat{x}_0 | \phi_0 \rangle ,
\end{eqnarray}
because $\epsilon(x_0) \ge | \langle \phi_0,\xi_0 | (\hat{x}_0)_{\rm exp} | \phi_0,\xi_0 \rangle - \langle \phi_0| \hat{x}_0 | \phi_0 \rangle |$.
However, the average value of $(\hat{x}_0)_{\rm exp} ^{\rm Ozawa}$ does not generally coincide with that of the object position $\hat{x}_0$ except for the special case,
\begin{eqnarray}
   \beta_1=1, \quad \beta_2=0.
\end{eqnarray}
This is the case where Ozawa's measurement result $(\hat{x}_0)_{\rm exp} ^{\rm Ozawa}$ is identical to the correct result $(\hat{x}_0)_{\rm exp}$ given by Eq. (3$\cdot$2).

Furthermore, it is easily found that his measurement result $(x_0)_{\rm exp} ^{\rm Ozawa}$ predicts an incorrect measurement result except for the special case with parameters (4$\cdot$4), when the initial state of the system is the eigenstate $|x_0,X_0 \rangle$ of $\hat{x}_0$ and $\hat{X}_0$.
(Strictly speaking, we must adopt wave packets with very small standard deviation instead of $|x_0 \rangle$ and $|X_0 \rangle$, because they are unnormalizable.)
For example, for the interaction with $\beta_1=1/2$ and $\beta_2=1/2$, $({x}_0)_{\rm exp} ^{\rm Ozawa}$ is equal to $x_0/2 + X_0/2$, the value of which is not generally identical to the true value $x_0$.
Thus, the measurement result defined by Ozawa does not reproduce correct measurement results except for the special case.

Furthermore, for interaction (2$\cdot$7), we obtain 
\begin{eqnarray}
   \epsilon^{\rm Ozawa}(x_0) = (\beta_1-1)^2 \sigma(x_0)^2+\beta_2^2\sigma(X_0)^2+( \langle \phi_0,\xi_0 | (\hat{x}_0)_{\rm exp} ^{\rm Ozawa}| \phi_0,\xi_0 \rangle - \langle \phi_0| \hat{x}_0 | \phi_0 \rangle )^2 .  \nonumber
\end{eqnarray}
This indicates that for interaction (2$\cdot$7) with $\beta_1 \ne 1$, Ozawa' theory does not reproduce the Born rule, because in this case Ozawa's measurement error $\epsilon^{\rm Ozawa}(x_0)$ does not become zero for any $| \phi_0 \rangle $ with $\sigma(x_0) \ne 0$ . The Born rule should be reproduced for any $| \phi_0 \rangle $.

These facts indicate that Ozawa's measurement error (4$\cdot$1) is not the error of the measurement result obtained for the pre-measurement position $\hat{x}_0$. It seems to express the noise caused when the information about the object position is transferred into the probe.
Therefore, when $\epsilon^{\rm Ozawa}(x_0)=0$, we can obtain the complete information about the initial object position $\hat{x}_0$ by measuring the probe position $\hat{X}_t$ just after the measurement.
However, Ozawa's error defined in such a manner is not the error of the measurement result.

Ozawa also defined the measurement error $\epsilon^{\rm Ozawa}(x_t)$ for the post-measurement position $\hat{x}_t$:\cite{MOPRL}
\begin{eqnarray}
   \epsilon^{\rm Ozawa}(x_t) &=& \langle \phi_0,\xi_0 | \{ (\hat{x}_t)_{\rm exp}^{\rm Ozawa}-\hat{x}_t \} ^{2}| \phi_0,\xi_0 \rangle ^{1/2}, \\
 (\hat{x}_t)_{\rm exp}^{\rm Ozawa} &\equiv& \hat{X}_t.
\end{eqnarray}
It was named ``resolution'' in Ref.~\citen{MOPRL}.

In the case of $\epsilon^{\rm Ozawa}(x_t)=0$, the average value of $(\hat{x}_t)_{\rm exp}$ should be identical to the average position of the object just after the measurement for any $| \phi_0 \rangle$ in the same way as Eq. (4$\cdot$3):
\begin{eqnarray}
   \langle \phi_0,\xi_0 | (\hat{x}_t)_{\rm exp} | \phi_0,\xi_0 \rangle =    \langle \phi_0,\xi_0 | \hat{x}_t | \phi_0,\xi_0 \rangle . 
\end{eqnarray}
Then, we can obtain the relations
\begin{eqnarray}
   \beta_1=\alpha_1, \quad \beta_2=\alpha_2,
\end{eqnarray}
for the measurement result $(\hat{x}_0)_{\rm exp}^{\rm Ozawa}$ .
From Eqs. (2$\cdot$8) and (4$\cdot$8), we obtain $1/\Gamma=\alpha_1\beta_2-\alpha_2\beta_1=0$, which does not satisfy the condition $|\Gamma|=1$ that the evolution $\hat{U}$ is unitary.
This fact shows that there exists no unitary evolution $\hat{U}$ that conserves the total momentum, when Ozawa' measurement result (4$\cdot$6) is adopted in the case of $\epsilon^{\rm Ozawa}(x_t)=0$.

Contrary to these facts, our measurement results (3$\cdot$2) and (3$\cdot$6) satisfy automatically average conditions (4$\cdot$3) and (4$\cdot$7), respectively, for any unitary evolution determined by Eq. (2$\cdot$7).
Furthermore, because our measurement error $\epsilon(x_0)$ (3$\cdot$3) does not depend on the initial state $| \phi_0 \rangle $ of the object, our theory does produce the Born rule for any $| \phi_0 \rangle $.

\section{Formulation of momentum disturbance}

We now derive the expression for the momentum disturbance.
Using Eq. (2$\cdot$7), we obtain
\begin{eqnarray}
   \hat{U}|x_0,X_0 \rangle = |\alpha_1 x_0+\alpha_2 X_0, \beta_1 x_0 + \beta_2 X_0 \rangle,
\end{eqnarray}
 for an arbitrary eigenket $|x_0 ,X_0 \rangle $. Then, we have
\begin{eqnarray}
   \hat{U} | p_0, P_0 \rangle = \int \hat{U} | x_0, X_0 \rangle \langle x_0 | p_0 \rangle \langle X_0 | P_0 \rangle  dx_0dX_0.  \nonumber
\end{eqnarray}
Consequently, we obtain
\begin{eqnarray}
   \langle p, P | \hat{U} | p_0, P_0 \rangle &=& \delta(p_0- \alpha_1 p-\beta_1 P) \delta(P_0- \alpha_2 p-\beta_2 P).
\end{eqnarray}
From the above equation, the object and probe momenta after the measurement are 
\begin{eqnarray}
   \hat{p}_t &=& a_1\hat{p}_0+a_2\hat{P}_0,  \\
   \hat{P}_t &=& b_1\hat{p}_0+b_2\hat{P}_0,
\end{eqnarray}
where $\hat{p}_0$ and $\hat{P}_0$ are the object and probe momenta before the measurement, respectively, and
\begin{eqnarray}
   a_1= \Gamma \beta_2, \quad a_2= -\Gamma \beta_1, \quad b_1= -\Gamma \alpha_2, \quad b_2= \Gamma \alpha_1.
\end{eqnarray}

When the initial states of the object and probe are eigenstates $|p_0 \rangle$ and $|P_0 \rangle$ of the momentum operators $\hat{p}_0$ and $\hat{P}_0$, respectively, by using Eq. (5$\cdot$3), the square of the momentum disturbance in the object caused by the position measurement is $\{(a_1 p_0+ a_2 P_0) -p_0 \}^2$.
Therefore, when the initial states of the object and probe are $|\phi_0 \rangle$ and $|\xi_0 \rangle$, respectively, the square of the momentum disturbance $(\Delta p)_{\rm dis}$ is
\begin{eqnarray}
(\Delta p)_{\rm dis}^2 &=& \int \{(a_1p_0+a_2P_0)-p_0 \}^2 | \langle p_0 | \phi_0 \rangle |^2 | \langle P_0 | \xi_0 \rangle |^2 dp_0dP_0 \nonumber  \\
               &=& \langle \phi_0,\xi_0|(\hat{p}_t-\hat{p}_0)^2 |\phi_0,\xi_0 \rangle.\end{eqnarray}
This is in agreement with the formulation given by Ozawa.\cite{MOPL1,MOPL2}

\section{Uncertainty relation between position uncertainty and momentum disturbance}

In $\S$3, we have derived two types of position-measurement errors $\epsilon(x_0)$ and $\epsilon(x_t)$.
In addition to these uncertainties, the standard deviations of the initial position $\hat{x}_0$ and experimental result $(\hat{x}_0)_{\rm exp}$ and others can also be regarded as the position uncertainty of the object.
Which uncertainty should be adopted as the position uncertainty in uncertainty relation (1$\cdot$2)?

Heisenberg illustrated three examples of position measurement in his text:\cite{WHPP} \ the $\gamma$-ray microscope, an electron passing through a slit of width $d$, and a Wilson chamber.
In these examples, he discussed the uncertainty in the measurement result for the post-measurement position $\hat{x}_t$, and he concluded that the uncertainty relation is valid for \textit{the object motion after the measurement}.
This indicates that the Heisenberg uncertainty relation given by Eq. (1$\cdot$2) should be interpreted as the relation between the measurement error $\epsilon(x_t)$ for the post-measurement position $\hat{x}_t$ and the momentum disturbance, 
because the uncertainty in the object position after the measurement is determined by the measurement error $\epsilon(x_t)$.
Heisenberg did not distinguish between the measurement errors $\epsilon(x_t)$ and $\epsilon(x_0)$.
This might be due to his implicit assumption that the object positions before and after the measurement are identical.
In this case, both measurement errors $\epsilon(x_t)$ and $\epsilon(x_0)$ are identical in our position-measurement model.

We now derive the uncertainty relation between the uncertainty in the 
position-measurement result and the momentum disturbance. 
As is well known, for any two observables $\hat{A}$ and $\hat{B}$, we obtain
\begin{eqnarray}
 \sigma(A) \sigma(B) \geq \frac{1}{2} | \langle [ \; \hat{A}, \; \hat{B} \; ] \rangle |, \nonumber
\end{eqnarray}
where $\sigma(A)$ and $\sigma(B)$ are the standard deviations of $\hat{A}$ and $\hat{B}$, respectively.\cite{MESSIAH} \ 
From the definition of the standard deviation, we obtain
\begin{eqnarray}
 \langle \hat{A}^2 \rangle = \sigma^2(A) + \langle \hat{A} \rangle^2 \geq  \sigma^2(A).  \nonumber
\end{eqnarray}
Therefore, the following equations also hold:
\begin{eqnarray}
 \sigma(A) (\langle \hat{B}^2 \rangle )^{1/2} &\geq& \frac{1}{2} | \langle [ \; \hat{A}, \; \hat{B} \; ] \rangle |,  \\
 (\langle \hat{A}^2 \rangle)^{1/2} (\langle \hat{B}^2 \rangle)^{1/2} &\geq& \frac{1}{2} | \langle [ \; \hat{A}, \; \hat{B} \; ] \rangle |.
\end{eqnarray}

Using the above inequality (6$\cdot$2) and the following commutation relation 
\begin{eqnarray}
[\; (\hat{x}_t)_{\rm exp} - \hat{x} _t , \; \hat{p}_t - \hat{p}_0 \; ] = -i\hbar,
\end{eqnarray}
we obtain 
\begin{eqnarray}
   \epsilon(x_t) (\Delta p)_{\rm dis} &\geq& \hbar/2.
\end{eqnarray}
This indicates that the uncertainty relation holds when the measurement error $\epsilon(x_t)$ for the post-measurement position is adopted as the position uncertainty.
It should be noted that uncertainty relation (6$\cdot$4) holds also for interaction (2$\cdot$3) proposed by Ozawa to demonstrate that there exists a measurement that violates uncertainty relation (1$\cdot$2).
This type of uncertainty relation is considered to be the original one intended by Hisenberg.

However, when the measurement error $\epsilon(x_0)$ for the pre-measurement position is adopted, the relation does not hold except for $|\beta_2|=1$:
\begin{eqnarray}
   \epsilon(x_0) (\Delta p)_{\rm dis} &\geq& |\beta_2|\hbar/2, 
\end{eqnarray}
since its commutation relation is 
\begin{eqnarray}
[ \; ( \hat{x}_0 )_{\rm exp} - \hat{x}_0 , \; \hat{p}_t - \hat{p}_0 \; ] &=& - \Gamma\beta_2(i \hbar).
\end{eqnarray}

The reason for this is apparent from the following expression of the standard deviation $\sigma((x_0)_{\rm exp})$ of the measurement result $(x_0)_{\rm exp}$:
\begin{eqnarray}
   \sigma((x_0)_{\rm exp}) &=& \{ \langle \phi_0,\xi_0 | \{ (\hat{x}_0)_{\rm exp} - \langle (\hat{x}_0)_{\rm exp} \rangle \} ^{2}| \phi_0,\xi_0 \rangle \}^{1/2},  \nonumber  \\
              &=& \{ (\sigma(x_0))^2 + (\epsilon(x_0))^2 \}^{1/2}.
\end{eqnarray}
This indicates that when one obtains a certain measurement result $(x_0)_{\rm exp}$ for the pre-measurement position by a one-time measurement, the uncertainty in its result is not the measurement error $\epsilon(x_0)$.
Even if the measurement error $\epsilon(x_0)$ is $0$, the measurement result $(x_0)_{\rm exp}$ has an uncertainty of the magnitude of the standard deviation $\sigma(x_0)$.
Only when $\sigma(x_0)$ is $0$, its uncertainty is the measurement error $\epsilon(x_0)$. However, when $\sigma(x_0)$ is not $0$, its uncertainty cannot be considered to be the measurement error $\epsilon(x_0)$ because we do not know which position component of the initial object state $|\phi_0 \rangle $ is measured in the one-time measurement.
In contrast to the uncertainty in the measurement result $(x_0)_{\rm exp}$, that in the measurement result $(x_t)_{\rm exp}$ for the post-measurement position $\hat{x}_t$ is unaffected by the initial uncertainty $\sigma(x_0)$.

When the object positions before and after the measurement are equal, we obtain $\alpha_1=1$ and $\alpha_2=0$ from Eq. (2$\cdot$7). In this case, the uncertainty relation given by Eq. (6$\cdot$5) holds because $| \beta_2 |=1$.

As mentioned above, the uncertainty in the measurement result $(x_0)_{\rm exp}$ for the pre-measurement position is not the measurement error $\epsilon(x_0)$ but the standard deviation $\sigma((x_0)_{\rm exp})$ of the measurement result $(x_0)_{\rm exp}$.
Using the inequality (6$\cdot$1) and the following commutation relation 
\begin{eqnarray}
 [\; (\hat{x}_0)_{\rm exp}            , \; \hat{p}_t - \hat{p}_0 \; ] &=& -i\hbar, 
\end{eqnarray}
we obtain 
\begin{eqnarray}
   \sigma((x_0)_{\rm exp}) (\Delta p)_{\rm dis}   &\geq& \hbar/2.
\end{eqnarray}
Thus, the uncertainty relation holds when the standard deviation $\sigma((x_0)_{\rm exp})$ of the measurement result $(x_0)_{\rm exp}$ is adopted as the position uncertainty.
This result is very reasonable because the uncertainty in the measurement result $(x_0)_{\rm exp}$ is the standard deviation $\sigma((x_0)_{\rm exp})$ and not the measurement error $\epsilon(x_0)$.

When the initial uncertainty $\sigma(x_0)$ of the object position is significantly smaller than the measurement error $\epsilon(x_0)$, the uncertainty relation for the measurement error $\epsilon(x_0)$ can be considered to be valid since in this case, $\sigma((x_0)_{\rm exp}) \simeq \epsilon(x_0)$.

As is well known, the quantum measurement process causes the reduction of a wave packet, by which the object state, initially in a superposition of different eigenstates, reduces to a single one of the states, and all other terms in the superposition have vanished after the measurement from an observer's view.
Thus, the object state after the measurement becomes unrelated to the initial uncertainty $\sigma(x_0)$ of the object position.
In the case where the measurement error $\epsilon(x_0)$ is not zero, a similar phenomenon occurs, although the problem is a little more complicated.
It is found in this case that the position measurement for obtaining the measurement result $(x_0)_{\rm exp}$ by the readout value $X$ of the probe position $\hat{X}_t$ filters the object position components $x_0$ in a region of order $\epsilon_X(x_0)$ about the point $(x_0)_{\rm exp}$ and all the other components have vanished after the measurement.
Therefore, the uncertainty in the measurement result for the post-measurement position $\hat{x}_t$ does not include the initial uncertainty $\sigma(x_0)$. It can be considered to be the measurement error $\epsilon(x_t)$.
In fact, it is proven in our measurement model that the standard deviation of position in the post-measurement state $| \phi_t \rangle _X$ of the object, where the subscript $X$ indicates that it is obtained when the readout value of the position $\hat{X}_t$ is $X$, is less than the measurement error $\epsilon_X(x_t)$ in Eq. (3$\cdot$10).

Contrary to this, in general, the position $\hat{x}_0$ has no definite value before the measurement.
When the measurement result $(x_0)_{\rm exp}$ is obtained for the position $\hat{x}_0$ with the error $\epsilon_X(x_0)$, it does not mean that it has the value in the region of order $\epsilon_X(x_0)$ about the point $(x_0)_{\rm exp}$ before the measurement.
Therefore, the uncertainty in the result for the pre-measurement position $\hat{x}_0$ should be considered to be the standard deviation $\sigma((x_0)_{\rm exp})$ of the measurement result $(x_0)_{\rm exp}$, not the measurement error $\epsilon(x_0)$.
This difference is caused by the reduction of the wave packet that occurs after the measurement.

As shown in Eqs. (3$\cdot$12) and (3$\cdot$13), the measurement errors $\epsilon(x_0)^2$ and $\epsilon(x_t)^2$ are the averages of $\epsilon_X(x_0)^2$ and $\epsilon_X(x_t)^2$, respectively, over the possible measurement results of the probe position $\hat{X}_t$. One might then arrive at the question of whether the following uncertainty relation for an \textit{individual} measurement process holds:
\begin{eqnarray}
   \epsilon_X(x_t) (\Delta p)_{\mbox{dis},X} &\geq& \hbar/2.
\end{eqnarray}
Similar uncertainty relations can also be considered for relations (6$\cdot$5) and (6$\cdot$9).
This problem is now under investigation.

\section{Justification for the linearity assumption}

We have proven under linearity assumption (2$\cdot$7) that Heisenberg uncertainty relations (6$\cdot$4) and (6$\cdot$9) hold.
We have also presented interaction (2$\cdot$5) that leads to linearity relation (2$\cdot$7) and conserves the total momentum of the object and probe.
We now examine the validity of the linearity assumption.

It is assumed in our position-measurement model that after the interaction of the object with the probe, the probe position $\hat{X}_t$ is measured, and by using this value, the measurement results of the object positions $\hat{x}_0$ and $\hat{x}_t$ are estimated.
When the initial state of the system composed of the object and probe is $| x_0, X_0 \rangle $, by generalizing Eq. (5$\cdot$1), the state after the interaction is given by
\begin{eqnarray}
   \hat{U}| x_0, X_0 \rangle = | f(x_0, X_0), g(x_0, X_0) \rangle ,
\end{eqnarray}
where $f(x_0, X_0)$ and $g(x_0, X_0)$ are arbitrary functions of $x_0$ and $X_0$.

When the right-hand side of Eq. (7$\cdot$1) is a superposition of more than two states of the system, we cannot determine the measurement results of the object positions uniquely by the readout of the probe position $\hat{X}_t$. 
Therefore, the right-hand side of Eq. (7$\cdot$1) must have only one term. 
When we write $f(x_0, X_0)$ and $g(x_0, X_0)$ as 
\begin{eqnarray}
   x_t = f(x_0, X_0),   \\
   X_t = g(x_0, X_0),
\end{eqnarray}
then it is easily found from Eq. (7$\cdot$1) that
\begin{eqnarray}
   \hat{x}_t \equiv \hat{U}^{-1} \hat{x}_0 \hat{U} = f(\hat{x}_0, \hat{X}_0),   \\
   \hat{X}_t \equiv \hat{U}^{-1} \hat{X}_0 \hat{U} = g(\hat{x}_0, \hat{X}_0).
\end{eqnarray}

von Neumann and Ozawa derived position operators (2$\cdot$2) and (2$\cdot$4) after the measurement using Hamiltonians (2$\cdot$1) and (2$\cdot$3), respectively, and neglecting the free Hamiltonians of the object and probe.
It should be noted that these results given by Eqs. (2$\cdot$2) and (2$\cdot$4) are only approximate ones.
In the same manner, we presented position operators (2$\cdot$6) after the measurement from  Hamiltonian (2$\cdot$5) that conserves the total momentum of the object and probe.
However, Eq. (7$\cdot$1) defines the evolution operator $\hat{U}$ directly without using a Hamiltonian.
Therefore, the position operators $\hat{x}_t$ and $\hat{X}_t$ derived in this section are free from the approximation of neglecting the free Hamiltonians of the object and probe.

The law of conservation of the linear momentum leads to the equation
\begin{eqnarray}
  \frac{\partial g}{\partial x_0} +\frac{\partial g}{\partial X_0}=1 .   \nonumber
\end{eqnarray}
A general solution for the above equation is 
\begin{eqnarray}
  X_t = g(x_0, X_0) = X_0 + G(x_0-X_0),
\end{eqnarray}
where $G(x)$ is an arbitrary function.
Because the value of $x_0$ must be determined uniquely by the value of $X_t$, the real number $X_t$ must have a one-to-one correspondence to $x_0$. Then, there exists an inverse function $G^{-1}(x)$. Thus, we obtain
\begin{eqnarray}
  x_0 = X_0 + G^{-1} (X_t-X_0).    \nonumber
\end{eqnarray}

The errors of position-measurement results are caused by the spreading of the probability density for the probe.
We cannot know which position component $X_0$ of the probe wave funcion interacts with the object.
Therefore, we define the measurement result $(x_0)_{\rm exp}$ for the pre-measurement position $x_0$ by the value of $x_0$ obtained when $X_0$ is set equal to $\langle \hat{X}_0 \rangle $ in the above equation:
\begin{eqnarray}
  (x_0)_{\rm exp} = \langle \hat{X}_0 \rangle + G^{-1} (X_t-\langle \hat{X}_0 \rangle).
\end{eqnarray}
Note that $X_t$ in the above equation is given by Eq. (7$\cdot$6).

Regarding $(x_0)_{\rm exp}$ as a function of $x_0$ and $X_0$, we expand $(x_0)_{\rm exp}$ into power series:
\begin{eqnarray}
  (x_0)_{\rm exp} = C_0+C_1(X_0-\langle \hat{X}_0 \rangle)+C_2(X_0-\langle \hat{X}_0 \rangle)^2+\ldots ,  
\end{eqnarray}
where $C_i$ represents functions of $x_0$ and $\langle \hat{X}_0 \rangle $, and clearly, $C_0=x_0$.

As discussed in $\S$4, when the measurement error $\epsilon(x_0)$ defined by Eq. (3$\cdot$3) is zero, the following condition must be satisfied for any possible $| \phi_0 \rangle $:\begin{eqnarray}
\langle \phi_0,\xi_0|(\hat{x}_0)_{\rm exp} |\phi_0,\xi_0 \rangle = \langle \phi_0|(\hat{x}_0)|\phi_0 \rangle .
\end{eqnarray}
Furthermore, we require condition (7$\cdot$9) to be satisfied also for any measurement error that is not zero.
Since the object cannot be regarded as being a specified state, clearly, condition (7$\cdot$9) must be satisfied for any $|\phi_0 \rangle $.
As for the probe, it is set to be a prescribed state before the measurement.
As suggested in Eq. (3$\cdot$3), in order to change the error $\epsilon(x_0)$, the probability density $|\xi_0(X_0)|^2$ of the probe must be changed.
Thus, condition (7$\cdot$9) must be satisfied for the changed probability density.
The above discussion indicates that condition (7$\cdot$9) must be satisfied for any $| \phi_0 \rangle $ and $|\xi_0 \rangle$.
Then, in the position representation, we obtain
\begin{eqnarray}
x_0 = \int ({x}_0)_{\rm exp} |\xi_0(X_0)|^2 dX_0.     \nonumber
\end{eqnarray}
The condition that the above equation holds for an arbitrary $x_0$ leads to the condition
\begin{eqnarray}
\int h(x_0, X_0) |\xi_0(X_0)|^2 dX_0 = 0,   
\end{eqnarray}
where
\begin{eqnarray}
h(x_0, X_0) \equiv  C_2(X_0-\langle \hat{X}_0 \rangle )^2 + C_3(X_0-\langle \hat{X}_0 \rangle)^3+\ldots .   \nonumber
\end{eqnarray}
Since Eq. (7$\cdot$10) must be satisfied for any probability density $|\xi_0(X_0)|^2$, we obtain
\begin{eqnarray}
h(x_0, X_0) = 0.     \nonumber
\end{eqnarray}
Consequently, we obtain
\begin{eqnarray}
  (x_0)_{\rm exp} = x_0+C_1(X_0-\langle \hat{X}_0 \rangle) .  
\end{eqnarray}

When in Eq. (7$\cdot$7) the function $G^{-1}(x)$ is expanded into power series
\begin{eqnarray}
  G^{-1}(x)=g_0+g_1x+g_2x^2+\ldots ,  \nonumber
\end{eqnarray}
where $g_i$ represents real numbers, the law of the conservation of parity leads to the relation $ g_0=0$.
Similarly, the function $G(x)$ is also expanded in power series in Eq. (7$\cdot$6).
If $X_t$ has terms higher than second order of $X_0$ in Eq. (7$\cdot$6), then it is found from Eq. (7$\cdot$7) that $(x_0)_{\rm exp}$ also has terms higher than second order of $X_0$.
Since $(x_0)_{\rm exp}$ in Eq. (7$\cdot$11) has no term higher than second order of $X_0$, $X_t$ has no term higher than second order of $X_0$.
Consequently, we obtain 
\begin{eqnarray}
  X_t = X_0+ \beta_1(x_0-X_0),
\end{eqnarray}
where $\beta_1$ is a real number.
This result is one of two relations given as the linearity assumption in Eq. (2$\cdot$7).

In the same way as Eq. (7$\cdot$6) was obtained, we obtain from Eq. (7$\cdot$2)
\begin{eqnarray}
  x_t = f(x_0, X_0) = x_0 + F(x_0-X_0),
\end{eqnarray}
where $F(x)$ is an arbitrary function.
We define the measurement result $(x_t)_{\rm exp}$ for the post-measurement position $x_t$  by the value of $x_t$ obtained when $X_0$ is set equal to $\langle \hat{X}_0 \rangle $ in Eq. (7$\cdot$13):
\begin{eqnarray}
  (x_t)_{\rm exp} = (x_0)_{\rm exp} + F((x_0)_{\rm exp} -\langle \hat{X}_0 \rangle).
\end{eqnarray}
Here, we used the fact that $x_0$ must be replaced by $(x_0)_{\rm exp}$ in this case.
In the same way as $(x_0)_{\rm exp}$, we expand $(x_t)_{\rm exp}$ into power series:
\begin{eqnarray}
  (x_t)_{\rm exp} = x_t+D_1(X_0-\langle \hat{X}_0 \rangle)+D_2(X_0-\langle \hat{X}_0 \rangle)^2+\ldots ,   \nonumber
\end{eqnarray}
where $D_i$ represents functions of $x_0$ and $\langle \hat{X}_0 \rangle $.

In the same way as the measurement result $(x_0)_{\rm exp}$ for the pre-measurement position, we require the following condition to be satisfied for any $|\phi_0 \rangle $ and $|\xi_0 \rangle $:
\begin{eqnarray}
\langle \phi_0,\xi_0|(\hat{x}_t)_{\rm exp} |\phi_0,\xi_0 \rangle = \langle \phi_0,\xi_0|\hat{x}_t |\phi_0,\xi_0 \rangle .
\end{eqnarray}
From the above condition, we obtain
\begin{eqnarray}
 \int i(x_0, X_0) |\xi_0(X_0)|^2 dX_0 = 0,   \nonumber   
\end{eqnarray}
where
\begin{eqnarray}
i(x_0, X_0) \equiv D_2(X_0-\langle \hat{X}_0 \rangle)^2+ D_3(X_0-\langle \hat{X}_0 \rangle)^3+\ldots .   \nonumber
\end{eqnarray}
For the same reason as we obtained $h(x_0, X_0)=0$, we have
\begin{eqnarray}
i(x_0, X_0) = 0.  \nonumber
\end{eqnarray}
Consequently, we obtain
\begin{eqnarray}
  (x_t)_{\rm exp} = x_t+D_1(X_0-\langle \hat{X}_0 \rangle) .
\end{eqnarray}
From Eqs. (7$\cdot$13) and (7$\cdot$14), we obtain
\begin{eqnarray}
  (x_t)_{\rm exp} -x_t = (x_0)_{\rm exp} -x_0 +F((x_0)_{\rm exp}-\langle \hat{X}_0 \rangle) - F(x_0-X_0).
\end{eqnarray}
From Eq. (7$\cdot$12), we obtain
\begin{eqnarray}
  (x_0)_{\rm exp} - x_0 = c_1(X_0-\langle \hat{X}_0 \rangle),
\end{eqnarray}
where $c_1$ is a real number and $c_1 \neq -1$ .
When in Eqs. (7$\cdot$13) and (7$\cdot$14) the function $F(x)$ is expanded into power series
\begin{eqnarray}
  F(x)=f_0+f_1x+f_2x^2+\ldots ,    \nonumber
\end{eqnarray}
where $f_i$ represents real numbers and $f_0=0$, it is easily found that $x_t$ has no term higher than second order of $X_0$.
Consequently, we obtain from Eq. (7$\cdot$13)
\begin{eqnarray}
  x_t = x_0 - \alpha_2(x_0-X_0),
\end{eqnarray}
where $\alpha_2$ is a real number. This is one of the two relations given as the linearity assumption.

Under the assumption that conditions (7$\cdot$9) and (7$\cdot$15) are satisfied for any $| \phi_0 \rangle $ and $|\xi_0 \rangle $, we have derived relations (7$\cdot$12) and (7$\cdot$19) that show that linearity assumption (2$\cdot$7) is valid.

There exist three elements ($\hat{U}$, $| \phi_0 \rangle $, $| \xi_0 \rangle$) that determine the measurement errors. The initial states $| \phi_0 \rangle $ and $| \xi_0 \rangle$ of the object and probe, respectively, are independent of each other.
The evolution operator $\hat{U}$ defined by Eq. (7$\cdot$1) does not depend on the states $| \phi_0 \rangle $ and $| \xi_0 \rangle$.
As discussed in $\S$4, Ozawa's theory does not reproduce the Born rule for any $| \phi_0 \rangle $ for interaction (2$\cdot$7) with $\beta_1 \ne 1$, because his measurement error (4$\cdot$1) depends on the initial state $| \phi_0 \rangle $.
Furthermore, if the measurement error depends on the state $| \phi_0 \rangle $, then it cannot be determined until the probability density $| \phi_0 (x_0) |^2$ can be estimated by repeating a measurement many times.
Therefore, it is reasonable to require that the measurement error $\epsilon(x_0)$ must be independent of $| \phi_0 \rangle $.
From Eq. (7$\cdot$8), we obtain
\begin{eqnarray}
  (x_0)_{\rm exp} -x_0 = C_1(X_0-\langle \hat{X}_0 \rangle)+C_2(X_0-\langle \hat{X}_0 \rangle)^2+\ldots .  \nonumber
\end{eqnarray}
Under the condition that the measurement error must be independent of $| \phi_0 \rangle $, the right-hand side of the above equation must not be a function of $x_0$.
Then, $(x_0)_{\rm exp}$ in Eq. (7$\cdot$8) does not have terms higher than second order of $x_0$.
Then, we can obtain relation (7$\cdot$11).
Similarly, the condition that the measurement error $\epsilon(x_t)$ must be independent of $| \phi_0 \rangle $ leads to relation (7$\cdot$16).
Therefore, by requiring that both measurement errors $\epsilon(x_0)$ and $\epsilon(x_t)$ are determined only by the evolution operator $\hat{U}$ and the initial state $| \xi_0 \rangle $ of the probe, and do not depend on $| \phi_0 \rangle $, we can justify linearity assumption (2$\cdot$7).

\section{Concluding remarks}

The Heisenberg uncertainty relation that is associated with the position-measurement process was examined under linearity assumption (2$\cdot$7) that the object position $\hat{x}_t$ and probe position $\hat{X}_t$ just after the measurement are expressed by the linear combination of the positions $\hat{x}_0$ and $\hat{X}_0$ just before the measurement.

The operators $(\hat{x}_0)_{\rm exp}$ and $(\hat{x}_t)_{\rm exp}$, which are those for the measurement results for the pre-measurement position $\hat{x}_0$ and post-measurement position $\hat{x}_t$, respectively, are defined by the relations
\begin{eqnarray}
   (\hat{x}_t)_{\rm exp} =\alpha_1 (\hat{x}_0)_{\rm exp} +\alpha_2 \langle \hat{X}_0 \rangle \hat{I}, \quad
   \hat{X}_t=\beta_1 (\hat{x}_0)_{\rm exp} + \beta_2 \langle \hat{X}_0 \rangle \hat{I}, \nonumber
\end{eqnarray}
which are obtained by the following replacements in Eq. (2$\cdot$7)
\begin{eqnarray}
   \hat{x}_0 \to (\hat{x}_0)_{\rm exp}, \quad
   \hat{x}_t \to (\hat{x}_t)_{\rm exp}, \quad
   \hat{X}_0 \to \langle \hat{X}_0 \rangle \hat{I} .   \nonumber
\end{eqnarray}

The average values of operators $(\hat{x}_0)_{\rm exp}$ and $(\hat{x}_t)_{\rm exp}$ obtained by the above method are identical to those of the object positions $\hat{x}_0$ and $\hat{x}_t$, respectively, for arbitrary $|\phi_0 \rangle $ and $|\xi_0 \rangle$, and for an arbitrary value of $\beta_1 \neq 0$.
There exists no position-measurement apparatus that has the evolution operator $\hat{U}$ with $\beta_1=0$.
Furthermore, we obtained the measurement errors $\epsilon(x_0)$ and $\epsilon(x_t)$ for the measurement results $(\hat{x}_0)_{\rm exp}$ and $(\hat{x}_t)_{\rm exp}$, respectively, on the basis of the physical consideration.
These errors are determined only by the evolution operator $\hat{U}$ and the initial state $| \xi_0 \rangle $ of the probe, and do not depend on the object state $| \phi_0 \rangle $.

Ozawa regarded the probe position $\hat{X}_t$ after the measurement as both the measurement results $(\hat{x}_0)_{\rm exp}$ and $(\hat{x}_t)_{\rm exp}$, i.e., $(\hat{x}_0)_{\rm exp}^{\rm Ozawa} = \hat{X}_t$ and $(\hat{x}_t)_{\rm exp}^{\rm Ozawa} = \hat{X}_t$.
It was shown that the measurement result $(\hat{x}_0)_{\rm exp}^{\rm Ozawa}$ does not predict generally correct measurement results except for the special case where Ozawa's measurement result $(\hat{x}_0)_{\rm exp} ^{\rm Ozawa}$ is identical to the correct result $(\hat{x}_0)_{\rm exp}$ given by Eq. (3$\cdot$2).
Ozawa's theory does not reproduce the Born rule for interaction (2$\cdot$7) with $\beta_1 \ne 1$, because his measurement error $\epsilon^{\rm Ozawa}(x_0)$ does not become zero for any $| \phi_0 \rangle $ in this case.
For interaction (2$\cdot$7), there exists no unitary evolution $\hat{U}$ that conserves the total momentum, when his measurement result $(\hat{x}_t)_{\rm exp}^{\rm Ozawa}$ is adopted and average condition (4$\cdot$7) is required to be satisfied.

We also derived the expression for the momentum disturbance, which is in agreement with the formulation given by Ozawa.

The uncertainty relation holds when the measurement error $\epsilon(x_t)$ for the object position $\hat{x}_t$ is adopted as the position uncertainty.
This relation is considered to be the original one that Heisenberg really intended.
The uncertainty relation also holds between the standard deviation $\sigma ((x_0)_{\rm exp})$ of the measurement result for the pre-measurement position $\hat{x}_0$ and the momentum disturbance.
These results seem to be reasonable, because the uncertainty in the measurement result $(x_0)_{\rm exp}$ for the pre-measurement position is the standard deviation $\sigma ((x_0)_{\rm exp})$, and not the measurement error $\epsilon(x_0)$, and because the uncertainty in the measurement result $(x_t)_{\rm exp}$ for the post-measurement position is the measurement error $\epsilon(x_t)$.

We have emphasized the importance of taking into account the conservation law of momentum.
The Wigner-Araki-Yanase theorem states that the presence of the conservation law limits the measurement of an observable that does not commute with the conserved quantity. \cite{EPW,AY,MY} \ The measurement of such an operator is only approximately possible.
However, in reality, one can measure the position with an arbitrarily small error, when the apparatus is sufficiently large, because the error decreases when the size of the apparatus increases. The Wigner-Araki-Yanase theorem does not put a limitation on the validity of the uncertainty relation between the position uncertainty and momentum disturbance.

The measurement model treated in this paper is an indirect measurement model \cite{MOAP} \ specified by $(\mathcal{K}, |\xi_0 \rangle, \hat{U}, \hat{M})$ consisting of a Hilbert space $\mathcal{K}$ of the probe, the initial state $|\xi_0 \rangle$ of the probe on $\mathcal{K}$, the time evolution operator $\hat{U}$, and an observable $\hat{M}$ on $\mathcal{K}$. Ozawa has showed that even though the indirect measurement models are only a subclass of all the possible quantum measurements, every measurement is statistically equivalent to one of indirect measurement models. \cite{MOAP} \

We have defined root-mean-square errors (3$\cdot$3) and (3$\cdot$5) using the indirect measurement model.
These amounts of errors apparently depend on the measurement model.
However, this is only apparently the case.
According to the study in Ref. ~\citen{MOAP}, errors (3$\cdot$3) and (3$\cdot$5) are given by the distance of POVM from an observable and are determined only by its POVMs, and hence, statistically equivalent apparatuses have the same amount of error.
For example, in our measurement model, a POVM $\Pi$ of an apparatus for measuring the observable $\hat{x}_0$ is given by 
\begin{eqnarray}
   \Pi(\Delta) &=& \mathrm{Tr}_\mathcal{K} \{ \hat{U}^{\dagger}(\hat{I} \otimes \hat{E}^M (\Delta)\hat{U}(\hat{I} \otimes | \xi_0 \rangle \langle \xi_0 |) \},    \\
   {\rm with} \quad \hat{M} &=& \frac{1}{\beta_1} \hat{X}_t - \frac{\beta_2}{\beta_1} \langle \hat{X}_0 \rangle \hat{I} ,  
\end{eqnarray}
using the partial trace operation $\mathrm{Tr}_\mathcal{K}$ over $\mathcal{K}$, where $E^M (\Delta)$ is the spectral projection of $\hat{M}$ corresponding to Borel set $\Delta$.

The validity of linearity assumption (2$\cdot$7) was examined in detail.
This assumption can be justified when the following conditions are satisfied:

(a) Measurement results $(x_0)_{\rm exp}$ and $(x_t)_{\rm exp}$ must be determined uniquely by the readout of the probe position $\hat{X}_t$ just after the measurement.

(b) The total momentum of the object and probe must be conserved before and after the measurement.

(c) The average values of $(\hat{x}_0)_{\rm exp}$ and $(\hat{x}_t)_{\rm exp}$ must be identical to those of $\hat{x}_0$ and $\hat{x}_t$, respectively, for arbitrary $|\phi_0 \rangle $ and $|\xi_0 \rangle $.

It was proven that the linearity assumption can also be justified when instead of condition (c) we require the condition that the measurement errors $\epsilon(x_0)$ and $\epsilon(x_t)$ are independent of the initial state $|\phi_0 \rangle $ of the object.

These are reasonable conditions to be required.

By using the results obtained in this paper, a standard quantum limit (SQL) for repeated measurements of free-mass position is proven to be valid.\cite{SK}
The problem of whether the SQL exists for monitoring free-mass position has been considered many times over the past years, in particular for gravitational-wave detection.

%


\begin{thebibliography}{99}

\bibitem{WHZP} W. Heisenberg, Z. Phys. \textbf{43} (1927), 172.
\bibitem{WHPP} W. Heisenberg, The Physical Principles of the Quantum Theory, University of Chicago Press, Chicago, 1930, reprinted by Dover, New York, 1949, 1967.
\bibitem{BrKhQM} V.B. Braginsky and F.Ya. Khalili, Quantum Measurement, Cambridge University Press, Cambridge, 1992.
\bibitem{MOPL1} M. Ozawa, Phys. Lett. A \textbf{299} (2002), 1.
\bibitem{MOPL2} M. Ozawa, Phys. Lett. A \textbf{318} (2003), 21.
\bibitem{MOPR} M. Ozawa, Phys. Rev. A \textbf{67} (2003), 042105.

\bibitem{KSPR} K. Koshino and A. Shimizu, Phys. Rep. \textbf{412} (2005), 191.
\bibitem{VNMF} J. von Neumann, Mathematical Foundations of Quantum Mechanics, Princeton University Press, Princeton, NJ, 1955.
\bibitem{Caves} C. M. Caves, Phys. Rev. Lett. \textbf{54} (1985), 2465.
\bibitem{MOPRL} M. Ozawa, Phys. Rev. Lett. \textbf{60} (1988), 385.
\bibitem{MESSIAH} A. Messiah, Quantum Mechanics, Dover Publications, New York, 1999.
\bibitem{EPW} E. P. Wigner, Z. Phys. \textbf{133} (1952), 101.
\bibitem{AY} H. Araki and M. M. Yanase, Phys. Rev. \textbf{120} (1960), 622.
\bibitem{MY} M. M. Yanase, Phys. Rev. \textbf{123} (1961), 666.
\bibitem{MOAP} M. Ozawa, Ann. Phys. \textbf{311} (2004), 350.

\bibitem{SK} S. Kosugi, to be published.


\end{thebibliography}
\end{document}